\documentstyle [11pt] {article}

\makeatletter

\newcommand{\singlespacing}{\let\CS=\@currsize\renewcommand{\baselinestretch}{1}\tiny\CS}
\newcommand{\oneandahalfspacing}{\let\CS=\@currsize\renewcommand{\baselinestretch}{1.25}\tiny\CS}
\newcommand{\doublespacing}{\let\CS=\@currsize\renewcommand{\baselinestretch}{1.35}\tiny\CS}

\def\@citex[#1]#2{\if@false\immediate\write\@auxout{\string\citation{#2}}\fi
  \def\@citea{}\@cite{\@for\@citeb:=#2\do
    {\@citea\def\@citea{,\linebreak[0]\hskip0pt plus .2em}%
      \@ifundefined{b@\@citeb}%
      {{\bf ?}\@warning{Citation `\@citeb' on page \thepage\space undefined}}%
      \hbox{\csname b@\@citeb\endcsname}}}{#1}}

\newtheorem{rule-def}[theorem]{Rule}

\begin{document}
\newcommand{\la}{\lambda}
\newcommand{\si}{\sigma}
\newcommand{\ol}{1-\lambda}
\newcommand{\be}{\begin{equation}}
\newcommand{\ee}{\end{equation}}
\newcommand{\bea}{\begin{eqnarray}}
\newcommand{\eea}{\end{eqnarray}}
\newcommand{\nn}{\nonumber}
\newcommand{\lb}{\label}

\begin{center}

{\large\bf A class of solutions for Einstein field equations with
spatially varying cosmological constant in spherically symmetric
anisotropic source}

\end{center}

\begin{center}

R. N. Tiwari$^{1}$ and Saibal Ray\footnote{E-mail:
saibal\_d\_ray@yahoo.co.in}$^{2,3}$
\\ $^{1}${\it Department of Mathematics, Indian Institute of
Technology, Kharagpur 721 302, W. B., India}\\ $^{2}${\it
Relativity and Cosmology Research Centre, Department of Physics,
Jadavpur University, Calcutta 700032, W. B., India}\\ $^{3}${ \it
Debra Higher Secondary School, Chakshyampur, Midnapore 721124, W.
B., India}\\

\end{center}

\vspace{1.0cm}

\abstract

In this work a class of interior solution for Einstein field
equations corresponding to a spherically symmetric anisotropic
fluid sphere has been obtained under the assumption that the
cosmological constant is spatially variable. The solution obtained
has the characteristics that the pressure and the cosmological
parameter vanish at the centre and at the boundary with a maximum
value somewhere inside the body. It has been argued that a
variable $ \Lambda $ is as much important physically in
Astrophysics as in Cosmology. \\

\vspace{.50cm}

1.~INTRODUCTION

The introduction of cosmological constant $\Lambda$ in Einstein
field equations has diverse opinions (Ellis$^{4}$,
Christopher$^{1,2}$, Curry$^{3}$). Theoretically, with the
introduction of $\Lambda$ the Einstein field equations become more
general than those without $\Lambda$. Observational evidences,
however, show that the contribution of $\Lambda$ is negligibly
small.

A variable $\Lambda$, on the other hand, has started attracting
the attention of theoreticians as well as experimentalists both
(Sakharov$^{9}$, Gunn and Tinslay$^{5}$, Peebles and Ratra$^{6}$,
Wampler and Burke$^{10}$, Ray and Ray$^{8}$).

Thus the present work is motivated to study the implications of
variable $\Lambda$ from astrophysical point of view. We consider
here an anisotropic massive static spherically symmetric fluid
core having Schwarzschild exterior field outside the boundary.

The scheme is as follows. In section 2,  field equations are
provided. Section 3 deals with the solutions whereas some comments
are are made in section 4. \\

2.~FIELD EQUATIONS\\

The Einstein field equations for the anisotropic fluid
distribution are given by \bea R_{ij} - \frac{1}{2}g_{ij} R +
g_{ij}\Lambda = -8\pi T_{ij}, \quad (G=c=1) \eea where the
matter-momentum tensor is given by ${T^{i}}_{j}=diag
(\rho,~-p_r,~-p_{\perp},-p_{\perp}$) and the related conservation
law here is (Peebles and Ratra$^{6}$) \bea {8\pi {T^{i}}_{j}};i=
-\Lambda;i \eea as the cosmological constant is assumed to be
spatially varying, i.e. $\Lambda=\Lambda(r)$.

We choose the spherically symmetric and static line element which
in the Schwarzschild coordinates ($t,~r,~\theta,~\phi) =
(0,~1,~2,~3)$ reads \bea ds^{2}= e^{\nu(r)} dt^{2} -
e^{\lambda(r)} dr^{2} - r^{2} ( d \theta ^{2} + sin^{2} \theta
d\phi^{2} ). \eea

Hence, the field equations (1) corresponding to the above line
element (3), are given by \bea e^{-\lambda} ( \lambda^{\prime}/r -
1/r^{2} ) + 1/r^{2} - \Lambda = 8\pi \rho, \eea \bea e^{-\lambda}
( \nu^{\prime}/r + 1/r^{2} ) - 1/r^{2} + \Lambda = 8\pi {p}_{r},
\eea \bea e^{-\lambda}[\nu^{{\prime}{\prime}}/2 +
{\nu^{\prime}}^{2}/4 - {\nu^{\prime} \lambda^{\prime}}/4 +
(\nu^{\prime} - \lambda^{\prime} )/ 2r] + \Lambda = 8\pi p_{\perp}
\eea where $p_r$ and $p_{\perp}$ are respectively the radial and
tangential pressures, and prime denotes derivative with respect to
radial coordinate $r$.

The conservation equation (2) becomes \bea \frac{d}{dr}(p_r -
{\Lambda}/{8 \pi} ) = - ( \rho + p_r ) {\nu^\prime}/2 + 2(
p_{\perp} - p_r)/r. \eea

Now, if we impose the condition \bea p_r = \frac{\Lambda}{8\pi}
\eea eqn. (5) yields \bea \nu^{\prime} = \frac{e^{\lambda} -
1}{r}. \eea

Substituting (8) and (9) in (4) and (6) we have \bea
(e^{\lambda}-1)(\rho + p_r) = 4(p_{\perp} - p_r). \eea

Also, from (7) by virtue of (8) we get \bea r \nu^{\prime} (\rho +
p_r) = 4(p_{\perp} - p_r). \eea

Again, eqn. (4) gives \bea e^{\lambda} = 1 - \frac{2M}{r} \eea
where \bea M = 4\pi \int^{r}_{0} r^2 (\rho + p_r) dr \eea which is
the Schwarzschild mass of a spherical body.\\

3.~A CLASS OF SOLUTIONS\\

To find out solutions of the field equations we assume that \bea
p_{\perp} = n p_{r},\quad(n \neq 1). \eea

Using (14), equation (10) reduces to \bea \rho = \frac{(4n
-3-e^{\lambda})p_r}{(e^{\lambda} - 1)} \eea which in turn makes
(13) to take the form \bea M = 16(n-1)\pi \int_{0}^{r}
\frac{r^2p_r}{e^{\lambda} - 1} dr. \eea

To solve the above integral, let us assume that \bea p_r =
k^2(e^{\lambda} - 1)(1 - r^2/a^2). \eea

This satisfies the physical conditions at the centre and at the
boundary, $a$ being the radius of the spherical distribution of an
isolated astrophysical system and $k^2$, a positive constant.

Using (17), we get the following solution:

\bea e^{-\lambda(r)} = 1 - AR^2 (5 - 3R^2), \eea

\bea e^{\nu(r)} = (1 - 2A)^{5/4} e^{\lambda(r)/4}
exp\left[\frac{5B}{2}\{tan^{-1}B(6R^2 - 5) - tan^{-1}B\}\right],
\eea

\bea \rho = \frac{15A(1-R^2)e^{\lambda}}{32(n-1)\pi a^2} [4(n - 1)
- (4n -3)AR^2 (5 - 3R^2)], \eea

\bea p_r = \frac{p_{\perp}}{n} = \frac{\Lambda}{8\pi} =
\frac{15A^2R^2(1-R^2)(5 - 3R^2)e^{\lambda}}{32(n-1)\pi a^2},
 \eea
and \bea M = \frac{AaR^3(5 - 3R^2)}{2}, \eea where $R = r/a$, $A =
32(n-1)\pi k^2 a^2/15$ and $B = [A/(12 - 25A)]^{1/2}$.

Matching of these solutions with the Schwarzschild exterior
solutions at the boundary $r=a$, yields \bea A = m/a, \quad B =
[m/(12a - 25m)]^{1/2}\eea where $m$ is the total active
gravitational mass.

The value of the constant $k$, occurring in eqn. (17), then can be
easily obtained by the relation \bea k^2 = 15m/[32(n-1)\pi a^3
\eea which is, obviously, always positive for $n >1$.\\

4. CONCLUSION\\

It may be seen from eqns. (10) and (11) that the case $n=1$ in
relation (14), i.e., when the pressure is isotropic, does not
exist in the present system as it makes the underlying space-time
flat. Hence, the assumption (8) leads only to anisotropic fluid.
The fluid density is constant at the centre with a value
$\rho(0)=15m/(8\pi a^3)$ and vanishes at the boundary. The
corresponding fluid pressure at $r=0$ is $p_r(0) =p_{\perp}(0)=0$
and $r=a$ is $p_r(a) =p_{\perp}(a)=0$. Since $p_r$ (or
$p_{\perp}$) is positive in the region $0 \leq r \leq a$, it
should have a maximum somewhere inside the fluid. For the present
case $p_r < p_{\perp}$ and hence the effective pressure acts
outwardly. Static equilibrium comes in the model by the adjustment
of this effective pressure to the gravitational pull of the matter
according to Ponce de Leon$^{7}$.

It is to be noted that like pressures the cosmological parameter
vanishes at the centre as well as at the boundary with a maximum
value somewhere inside the body. However, one can also have a
solution by changing the condition (8) to $p_r=(\Lambda -
\Lambda_0)/8 \pi$, where $\Lambda_0$ is a constant. The
cosmological parameter $\Lambda$ for this solution will not vanish
at the boundary. The exterior Schwarzschild solution for this case
will involve the erstwhile cosmological constant to which the
interior solution is to be matched. In the present solution the
scalar $\Lambda$ takes the role of compensatory balancing pressure
varying from point to point in exactly the same way as the radial
and the tangential pressures.\\

\begin{center}
 REFERENCES
\end{center}

\begin{enumerate}
\item{} R. Christopher, {\it Stud. Hist. Phil. Sci.} {\bf 21} (1990), 589.

\item{} R. Christopher, {\it Stud. Hist. Phil. Sci.} {\bf 23} (1992), 661.

\item{} C. Curry, {\it Stud. Hist. Phil. Sci.} {\bf 23} (199), 657.

\item{} G. F. R. Ellis, {\it Expanding Universe: A History of Cosmology from 1917 to 1960};
Einstein and the History of General Relativity, North Andover, MA
(1986), 367; Einstein Studies, Birkh{\"o}user Boston, MA (1988),
1.

\item{} J. Gunn  and B. M. Tinslay, {\it Nature} {\bf 257} (1975), 454.

\item{} P. J. E. Peebles and B. Ratra, {\it Astrophys. J.} {\bf 325} (1988),
417.

\item{} J. Ponce de Leon, {\it Gen. Rel. Grav.} {\bf 19} (1987), 197.

\item{} S. Ray and D. Ray, {\it Astrophys. Space Sci.} {\bf 203} (1993), 211.

\item{} A. D. Sakharov, {\it Doklady Akad. Nauk. SSSR } {\bf 177} (1968),
                70 (translated: {\it Soviet Phys. Doklady} {\bf 12}).

\item{} E. J. Wampler and W. L. Burke, {\it New Ideas in
                Astronomy} (Cambridge Univ. Press) (1988), p. 317.

\end{enumerate}

\end{document}